\documentclass[twocolumn,aps,prl,reprint,groupedaddress,showkeys]{revtex4-1}
\usepackage{graphicx,color}
\usepackage{color}

\newcommand{\be}{\begin{equation}}
\newcommand{\ee}{\end{equation}}

\begin{document}

\title{Enhanced thermionic-dominated photoresponse in graphene Schottky junctions}

\author{Joaquin F. Rodriguez-Nieva$^1$}
\author{Mildred S. Dresselhaus$^{1,2}$}
\author{Justin C. W. Song$^{3}$}
\affiliation{$^1$ Department of Physics, Massachusetts Institute of Technology, Cambridge, MA 02139 USA}
\affiliation{$^2$ Department of Electrical Engineering and Computer Science, Massachusetts Institute of Technology, Cambridge, MA 02139, USA}
\affiliation{$^3$ Walter Burke Institute for Theoretical Physics and Institute of Quantum Information and Matter, California Institute of Technology, Pasadena, CA 91125 USA}

\begin{abstract}
Vertical heterostructures of van der Waals materials enable new pathways to tune charge and energy transport characteristics in nanoscale systems. We propose that graphene Schottky junctions can host a special kind of photoresponse which is characterized by strongly coupled heat and charge flows that run vertically out of the graphene plane. This regime can be accessed when vertical energy transport mediated by thermionic emission of hot carriers overwhelms electron-lattice cooling as well as lateral diffusive energy transport. As such, the power pumped into the system is efficiently extracted across the entire graphene active area via thermionic emission of hot carriers into a semiconductor material. Experimental signatures of this regime include a large and tunable internal responsivity ${\cal R}$ with a non-monotonic temperature dependence. In particular, ${\cal R}$ peaks at electronic temperatures on the order of the Schottky potential $\phi$ and has a large upper limit ${\cal R} \le e/\phi$ ($e/\phi=10\,{\rm A/W}$ when $\phi = 100\,{\rm meV}$). Our proposal opens up new approaches for engineering the photoresponse in optically-active graphene heterostructures. 
\end{abstract}

\pacs{}
\keywords{graphene; hot carriers; Schottky junction; photocurrent}

\maketitle

Vertical heterostructures comprising layers of van der Waals (vdW) materials have recently emerged as a platform for designer electronic interfaces \cite{geimvdw}. Of special interest are heterostructures which feature tunable interlayer transport characteristics, as exemplified by g/X Schottky junctions \cite{gschottky2, gschottky1, gschottky3,gsolar,gsolar2,britnellscience,ws2heterostructures,heterostructures,britnellnanolett}; here `g' denotes graphene, and X is a semiconductor material, such as Si, MoS$_2$ or WSe$_2$. These junctions are characterized by Schottky barriers $\phi$ that span two orders of magnitude $\phi \approx 0.01 - 1 \, {\rm eV}$ and exhibit in situ control through applied bias or by using gate potentials \cite{ws2heterostructures,gschottky2, gschottky1,britnellscience,schottkyreport}. The wide range of $\phi$ achievable across the g/X interface, combined with the unique graphene photoresponse mediated by long-lived hot carriers (elevated electronic temperatures, $T_{\rm g}$, different from those of the lattice, $T_0$\cite{macdonald,wong,gabor,song,graham,betz}), make graphene Schottky junctions a prime target for accessing novel vertical energy transport regimes \cite{koppens}. 

\begin{figure} 
\centering \includegraphics[scale=1.0]{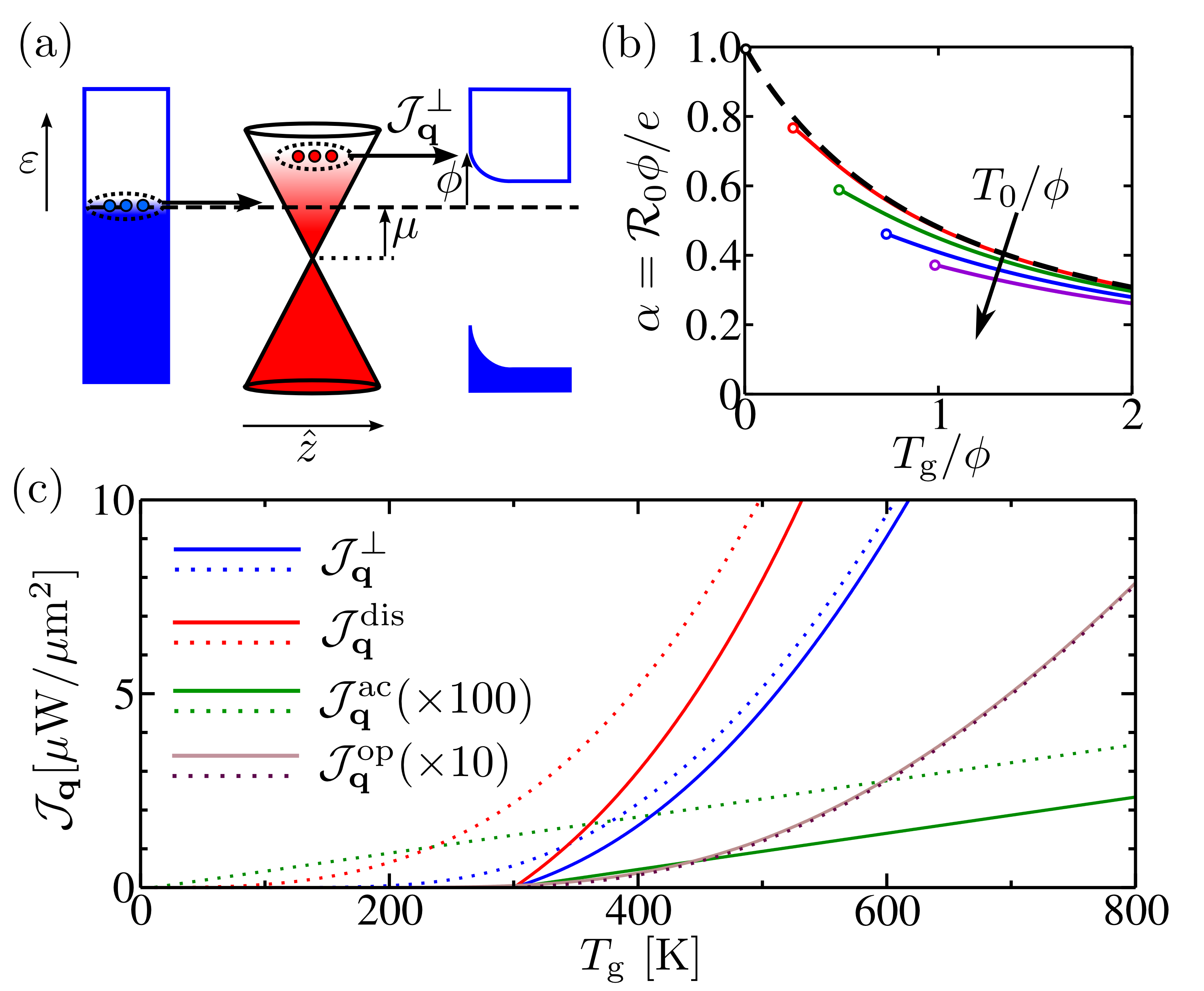}
\caption{
(a) Schematic of thermionic-dominated heat transport in a graphene Schottky junction: hot carriers with energies $\varepsilon$ close to the Schottky barrier height $\phi$ are thermionically emitted into a semiconductor material in the out-of-plane direction $\hat{z}$, while cold carriers are injected through an Ohmic contact at the Fermi level, $\mu$, generating a net vertical heat current ${\cal J}_{\bf q}^{\perp}$. (b) Normalized ideal responsivity, $\alpha = {\cal R}_0\phi/e$, is shown as a function of normalized graphene electronic temperature, $k_{\rm B} T_{\rm g}/\phi$, with $e$ the electron charge [see Eq.(\ref{eq:ideal})]. Curves are obtained for normalized ambient temperature $k_{\rm B} T_{0}/\phi=0,0.25,0.5,0.75,1.0$, indicated with different colors for increasing $T_0$; characteristic ${\cal R}_0 \approx e/\phi \sim 10\,{\rm A/W}$ can be large (for $\phi = 100\,{\rm meV}$, see text). (c) Thermionic cooling channel, ${\cal J}_{\bf q}^{\perp}$, compared with acoustic phonon cooling (clean case), ${\cal J}_{\bf q}^{\rm ac}$, disorder-assisted cooling, ${\cal J}_{\bf q}^{\rm dis}$, and optical phonon cooling ${\cal J}_{\bf q}^{\rm op}$ [see text and Eq.(\ref{eq:elatticecooling})], shown for ambient temperature $T_0 = 300\,{\rm K}$ (solid lines) and $T_0 = 0$ (dotted lines). Note that ${\cal J}_{\bf q}^{\perp}$ overwhelms ${\cal J}_{\bf q}^{\rm ac}$ and ${\cal J}_{\bf q}^{\rm op}$(clean case) and is competitive with ${\cal J}_{\bf q}^{\rm dis}$ (dirty case). Parameter values used: $\mu = \phi = 100\,{\rm meV}$, $k_{\rm F}\ell = 50$, and $G_0 = 10\,{\rm mS}/\mu{\rm m^2}$, see Eqs.\,(\ref{eq:currentadim}) and (\ref{eq:elatticecooling}). 
}
\label{fig:convcooling}
\end{figure}

Here we show that specially designed graphene Schottky junctions can host an enhanced thermionic-dominated photoresponse driven by strongly coupled charge and energy currents. Such photoresponse proceeds, as illustrated in Fig.\,\ref{fig:convcooling}a, via the thermionic emission of graphene hot carriers with energy larger than the Schottky barrier. At steady state, an equal number of cold carriers are injected at the Fermi surface through an ohmic contact, giving a net flow of heat ${\cal J}_{\bf q}^\perp$ out of the graphene electronic system balancing the energy pumped into the system. 

Strikingly, thermionic emission yields strong heat transport running vertically out of the hot electron system, which dominates over more conventional electronic cooling channels, e.g. electron-lattice cooling. Indeed, we find that ${\cal J}_{\bf q}^\perp$ can be significant in graphene (see Fig.\,\ref{fig:convcooling}c) when $k_{\rm B}T_{\rm g} \approx \phi/2$, dominating over acoustic and optical phonon cooling \cite{macdonald,wong} in pristine graphene Schottky junctions; ${\cal J}_{\bf q}^\perp$ also overwhelms in-plane (lateral) diffusive energy transport. We find that the values of ${\cal J}_{\bf q}^\perp$ are competitive with disorder-assisted cooling \cite{song,graham,betz} in more dirty devices. 

Graphene is essential to our proposal due to a unique combination of electronic characteristics. First, fast intraband Auger-type scattering \cite{klaas,songIE} allows the absorbed photon energy flux, ${\cal J}_{\bf q}^{\rm in}$, to be efficiently captured as heat by ambient carriers in graphene; this process results in a thermalized hot carrier distribution \cite{klaas,songIE}. Second, graphene is characterized by slow electron-lattice cooling mechanisms \cite{macdonald,wong,gabor,song,graham,betz} which enables $T_{\rm g} > T_0$ to drive a strong thermionic current. This is due to the large optical phonon energy in graphene \cite{macdonald,wong} as well as the weak electron-acoustic phonon coupling [for a detailed comparison between cooling rates, see Eq.(\ref{eq:elatticecooling}) below]. Third, the bias and gate-tunable work function allows an experimentally accessible way to optimize device operation, predicted to occur at $k_{\rm B} T_{\rm g} \approx \phi / 2$, for a range of technologically achievable temperatures and barrier materials. Indeed, whereas other Schottky junctions (e.g. Au/Si, Ag/Si) may also display vertical energy currents, their large Schottky barriers ($\phi \approx 1\, {\rm eV}$) and fast electron-lattice cooling render the thermionic-dominated regime impractical for these systems. 

\begin{figure}
\centering \includegraphics[scale=1.0]{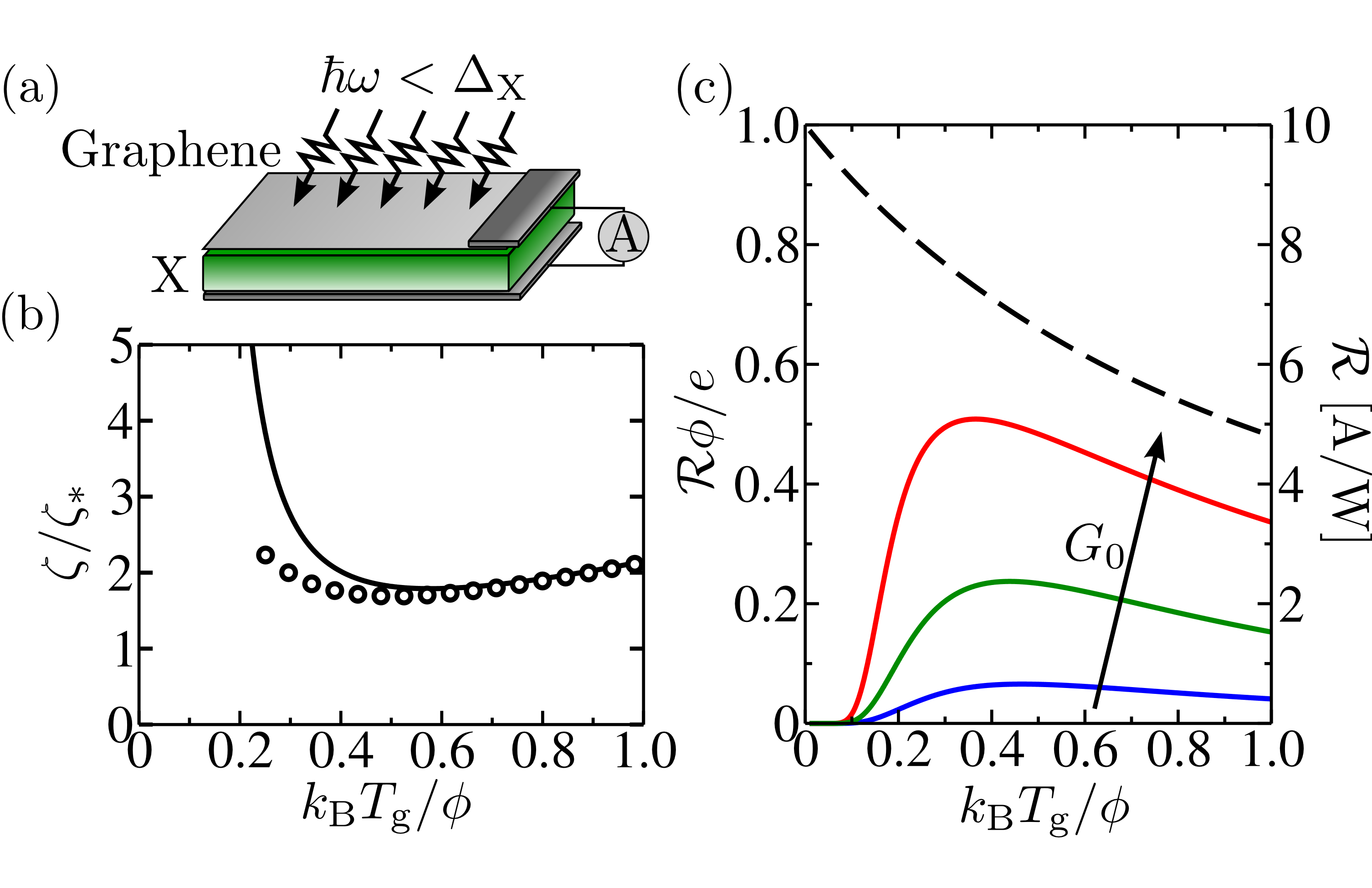}
\caption{
(a) Schematic of a hot carrier photodetector formed by a graphene-X Schottky junction, where X is a semiconductor material with bandgap $\Delta_{\rm X}$. Here we consider graphene as the photoactive material for absorption, i.e. photon energy $\hbar \omega < \Delta_{\rm X}$. 
(b) Ratio $\zeta = {\cal J}_{\bf q}^{\rm loss}/{\cal J}_{\bf q}^{\perp}$ modeled via Eq.(\ref{eq:kappa}) as a function of $T_{\rm g}$, with $T_0=0$ (solid line) and $T_0= 300 \, {\rm K}$ (empty circles). Note that the latter has a smaller range, $T_{\rm g} \geq T_0$.
(c) The responsivity ${\cal R}$ (solid lines) for the g/X junction exhibits a non-monotonic electronic temperature dependence peaking at $T_{\rm g} \approx \phi / 2k_{\rm B}$, shown for $\zeta$ modeled in Eq.(\ref{eq:kappa}) [panel (b)] with $\zeta_* = 5, 1, 0.2$ (blue, green, red, respectively); dimensionless ${\cal R}$ is shown on the left vertical axis. The dashed line indicates an ideal case ${\cal R} = {\cal R}_0$. Here we used values $G_0 = 2, 10,50 \,{\rm mS}/\mu{\rm m^2}$, $\mu = \phi = 100\,{\rm meV}$ and $k_{\rm F}\ell = 50$ yielding sizable ${\cal R}$ (right vertical axis). 
}
\label{fig:sens2}
\end{figure}

An important optoelectronic figure of merit is the conversion between incoming photon energy flux, ${\cal J}_{\bf q}^{\rm in}$, and detected photocurrent, ${\cal J}_e^\perp$, encoded in the (internal) responsivity ${\cal R} = {\cal J}_{e}^{\perp} / {\cal J}_{\bf q}^{\rm in}$. Importantly, our model yields a large upper limit for ${\cal R}$. Indeed, energy $\phi$ is transported per carrier extracted across the g/X junction (Fig.\,\ref{fig:convcooling}a) yielding a limiting internal responsivity ${\cal R}_0$, occurring in the thermionic-dominated regime (i.e., ${\cal J}_{\bf q}^{\perp} = {\cal J}_{\bf q}^{\rm in}$) given by 
\be
{\cal R} \le {\cal R}_0, \quad \quad {\cal R}_0= \frac{e}{\phi} \times\alpha(\tilde{T}_{\rm g},\tilde{T}_0).
\label{eq:ideal}
\ee
Here $e$ is the electron charge, $\tilde{T}_{{\rm g},0} = k_{\rm B} T_{{\rm g},0}/\phi$ are the dimensionless graphene and ambient temperatures (temperature of the lattice and semiconductor), respectively, and $\alpha$ is a dimensionless function (see text below) plotted in Fig.\,\ref{fig:convcooling}b. The function $\alpha$ can take values close to unity, allowing ${\cal R}_0$ to be on the order of $e/\phi = 10 \, {\rm A/W}$, for $\phi = 100 \, {\rm meV}$. For a discussion of net values of $\mathcal{R}$ in Eq.(\ref{eq:ideal}), see Fig.\ref{fig:sens2}c. 

Since the incident photon energy $\hbar \omega$ (e.g. in the visible) can be many multiples of $\phi$, we anticipate that g/X Schottky photodetectors can provide significant gains in the internal responsivity compared to those in conventional (photovoltaic-based) photodetectors, which are limited by $\mathcal{R}^{\rm PV} \le e/ \hbar \omega$ \cite{sze}. In particular, the ultra-fast electron energy relaxation times in graphene yield {\it multiple} hot carriers per absorbed photon \cite{klaas,songIE}, in stark contrast to photovoltaic-based schemes that yield a single electron-hole pair per absorbed photon. Naturally, the external responsivity of the device is also affected by the absorption coefficient of the photoactive material. Whereas the absorption coefficient of 2.3\% per layer in graphene \cite{gabsorption} is small compared, for instance, to typical values of 10-50\% in Si \cite{siabsorption}, this small value can be increased using optical waveguides \cite{gschottky3} and plasmon enhanced absoption \cite{fengnian}. These external enhancement mechanisms will not be discussed here. 

In addition, g/X photodetectors also enable a boosted photoresponse compared to previous photothermoelectric-based schemes \cite{gabor}. Indeed, the vertical structure allows to circumvent lateral electronic heat diffusion, which drastically reduces the operating electronic temperatures and efficiencies in photothermoelectric-based schemes. 

Another important feature of the g/X photoresponse is the possibility of using the temperature dependence of $\mathcal{R}$ as a diagnostic of thermionic-dominated photoresponse. Including losses to the lattice via disorder-assisted cooling, we find that ${\cal R}$ is non-monotonic, peaking at an optimal operating hot carrier temperature $k_{\rm B} T_{\rm g}\approx \phi / 2$ (Fig.\,\ref{fig:sens2}c). Since $T_{\rm g}$ can be controlled by the incident light power and $\phi$ via gate voltage, non-monotonic ${\cal R}$ as a function of ${T}_{\rm g}$ provides an easily accessible experimental signature of the strongly coupled charge and energy thermionic transport that is engineered across the g/X interface. Indeed, non-monotonic temperature behavior does not occur in the photovoltaic-based devices, where responsivity is mainly independent of pump power or photon intensity. 

We begin by modeling vertical transport across the g/X device, as depicted in Fig.\,\ref{fig:convcooling}a. To describe thermionic transport over the barrier, we adopt a quasielastic but momentum non-conserving approximation \cite{jrnnl}. This approximation is valid because, at high $T$, a number of momentum scattering mechanisms at the Schottky junction are possible, such as scattering by defects, intrinsic phonons and substrate phonons. Furthermore, the typical energy exchange in these processes is small on the Schottky barrier scale. As a result, we generically write the electron and heat current across the g/X Schottky junction as
\begin{eqnarray}
\left[ \begin{array}{c} {\cal J}_e^{\perp} \\ {\cal J}_{\bf q}^{\perp}\end{array} \right] &&= \displaystyle \int_{-\infty}^{\infty} d\varepsilon \left[ \begin{array}{c} e \\ \varepsilon \end{array}\right] G  (\varepsilon)
 \left[f\left(\frac{\varepsilon}{k_{\rm B}T_{\rm g}}\right)-f\left(\frac{\varepsilon}{k_{\rm B}T_0}\right)\right],  \nonumber \\
 G (\varepsilon)  &&= \displaystyle \frac{2\pi e^2}{\hbar} D_{\rm g}(\varepsilon)D_{\rm c}(\varepsilon)|{\rm T}(\varepsilon)|^2.
\label{eq:current}
\end{eqnarray}
Here $G(\varepsilon)$ is a parameter with units of electrical conductance which characterizes the Schottky interface (see discussion below), $D_{\rm g}$ ($D_{\rm c}$) is the density of states of graphene (the conduction band of the semiconductor), $f(x)=1/(e^x+1)$ is the Fermi distribution function, ${\rm T}(\varepsilon)$ is the energy-dependent tunneling transition matrix element between graphene and semiconductor electronic states, and energies $\varepsilon$ are referenced from the Fermi energy $\mu$ (see Fig.\,\ref{fig:convcooling}a). The function ${\rm T}(\varepsilon)$ contains all the microscopic information about the relevant mechanisms that couple graphene with material X, such as phonons or hot-spots formed by defects. 

Two important assumptions are present in Eq.(\ref{eq:current}). First, we neglected hole transport between graphene and the valence band of X assuming that the barrier height for hole transport is much larger than the corresponding one for electron transport. Secondly, we assume that the Fermi level and the temperature in graphene and X are spatially fixed. In a more realistic scenario, the pumping power may cause the temperature and Fermi level to spatially vary in the out-of plane direction. In this case, both quantities need to be determined self-consistently by appropriate balance equations. However, these do not introduce any new qualitative features to our simplified model. 

When light heats graphene electrons so that $T_{\rm g} > T_0$, Eq.(\ref{eq:current}) describes the short-circuit charge current (photocurrent) and the energy current flow, shown schematically in Fig.\,\ref{fig:convcooling}c. At steady state, $T_{\rm g}$ is determined by energy balance of the incident absorbed power in graphene, ${\cal J}_{\bf q}^{\rm in}$, and the energy being dissipated by the graphene electronic system, ${\cal J}_{\bf q}^{\rm out}$, that includes both the thermionic channel, ${\cal J}_{\bf q}^{\perp}$, and other dissipative channels, ${\cal J}_{\bf q}^{\rm loss}$ (e.g. electron-lattice cooling, and diffusive heat transport discussed below). Explicitly, we have 
\be
{\cal J}_{\bf q}^{\rm in} = {\cal J}_{\bf q}^{\rm out} = {\cal J}_{\bf q}^{\perp} (T_{\rm g}, T_0)+  {\cal J}_{\bf q}^{\rm loss} (T_{\rm g},T_0),
\label{eq:heatbalance}
\ee
where we have fixed $T_0$ to the temperature of the ambient environment, i.e. there is no backflow of hot electrons into graphene. The latter assumption results from the large heat capacity and fast electron-lattice cooling in highly doped semiconductors such as Si \cite{sicooling}. In what follows, we shall analyze the energy/charge characteristics of g/X Schottky junctions as a function of $T_{\rm g}$ and $T_0$; naturally the $T_{\rm g}$ values displayed can be attained via a suitably chosen ${\cal J}_{\bf q}^{\rm in}$. 

The depletion width, for example in g/Si interfaces\cite{gsolar2}, can be many times larger than the electron wavelength. As a result, only electrons with energies above the effective barrier $\phi$  formed at the g/X interface contribute to the current; in this way, the photocurrent is thermally activated. Here we adopt a phenomenological approach to capture the essential physics independent of the microscopic details of the device. To this end, we approximate $G(\varepsilon) = G_0 \Theta (\varepsilon -\phi)$ in Eq.(\ref{eq:current}), with $\Theta$ the step-function, in order to aggregate the microscopics of the junction into a single variable that can be easily measured in experiments. This approach does not describe field emission, which is considered negligible because we are limiting our discussion to the zero bias behavior, i.e. the closed circuit photocurrent. We emphasize that this approximation does not affect the qualitative behavior of ${\cal J}_e^{\perp}$, ${\cal J}_{\bf q}^{\perp}$ or $\alpha$ for the range of temperatures of interest, $T_{\rm g} \lesssim \phi$; further, this approximation represents a conservative estimate of the particle current, since $G(\varepsilon)$ is typically a monotonically increasing function due to the larger density of states available for scattering at larger $\varepsilon$ in graphene. As a side remark, we note that $G_0$ is {\it not} the zero-bias junction conductance; this quantity is suppressed by a factor $e^{-\phi/T}$ with respect to $G_0$, as discussed in the paragraph following Eq.(\ref{eq:kappa}). 

Using a step-like transmission, heat and charge currents can then be expressed in terms of non-dimensional integrals by defining $x  = \varepsilon/\phi$ in Eq.(\ref{eq:current}), yielding
\be
{\cal J}_e^{\perp} = \frac{G_0 \phi}{e}\int_{1}^{\infty}dx \, \Delta f (x)\,,\, {\cal J}_{\bf q}^{\perp} = \frac{G_0 \phi^2}{e^2}\int_{1}^{\infty}dx \, x \,\Delta f (x),
\label{eq:currentadim}
\ee
where $\Delta f(x) = f(x/\tilde{T}_{\rm g}) - f(x/\tilde{T}_0)$. It is straight-forward to show that the integrals on the right-hand side of Eq.(\ref{eq:currentadim}) are related to the well-known complete Fermi integrals, 
\be
F_k(\xi) = \int_{0}^{\infty}dx\, \frac{x^{k}}{e^{x-\xi}+1},
\label{eq:fermiint}
\ee
via $\int_{1}^{\infty}dx x^{n} f(x/\tilde{T}) = \sum_{k=0}^{n}\big( \begin{array}{c} n \\ k \end{array} \big)\tilde{T}^{k+1}F_{k}(-1/\tilde{T})$. In the low temperature regime, $\tilde{T} \ll 1$, the value of $F_k$ behaves as $F_{k}(-1/\tilde{T}) \approx \Gamma(k+1) e^{-1/\tilde{T}}$, with $\Gamma$ the Gamma function. In the high temperature regime, $\tilde{T} \gg 1$, $F_k$ takes values $F_0(-1/\tilde{T}) \approx \rm{ln}(2)$ and $F_1(-1/\tilde{T})\approx \pi^2/12$.

A key feature of thermionic-dominated [${\cal J}_{\bf q}^{\rm loss} = 0$] transport is the strong charge-energy current coupling manifested in ${\cal R}_0$. Using Eq.(\ref{eq:currentadim}) above, we obtain Eq.(\ref{eq:ideal}) with $\alpha$ given by
\be
\alpha (\tilde T_{\rm g}, \tilde T_0)= \frac{\int_1^\infty dx \Delta f(x) }{\int_1^\infty dx x \Delta f(x) }.
\label{eq:F}
\ee
The general behavior of $\alpha (\tilde T_{\rm g}, \tilde T_0)$ can be most easily understood by first setting $ T_0 = 0$. In this case, $\alpha$ in Eq.(\ref{eq:ideal}) adopts the simple form
\be
\alpha_0(\tilde{T}_{\rm g}) \equiv \alpha(\tilde{T}_{\rm g}, \tilde{T_0} =0) = \left[ 1 + \tilde{T}_{\rm g} \frac{F_{1}(-1/\tilde{T}_{\rm g})}{F_{0}(-1/\tilde{T}_{\rm g})}\right]^{-1}.
\label{eq:Fanalytic}
\ee
Importantly, $\alpha_0$ is a decreasing function of temperature $\tilde{T}_{\rm g}$, as shown in the black dashed curve of Fig.\,\ref{fig:convcooling}b. In particular, for $\tilde{T}_{\rm g} \ll 1$, $\alpha_0$ takes values $\alpha_0 \sim (1+\tilde{T}_{\rm g})^{-1}$ of order unity, and for $\tilde{T}_{\rm g}\gg 1$, $\alpha_0$ decreases with inverse temperature as $\alpha_0 \approx 12 {\rm log}(2) / (\pi^2 \tilde{T}_{\rm g})$ (see Fig.\,\ref{fig:convcooling}b). This latter fact means that, although ${\cal R}_0$ is expressed in units of $e/\phi$ in Eq.(\ref{eq:ideal}), ${\cal R}_0$ cannot grow indefinitely by making $\phi$ smaller; ${\cal R}_0$ reaches a saturating value ${\cal R}_0 \approx 12 {\rm log}(2) e / (\pi^2 k_{\rm B} T_{\rm g})$ for $T_{\rm g} \gg \phi$, as shown in Fig.\,\ref{fig:convcooling}c.

For finite values of $T_0$, the qualitative behavior of $\alpha$ does not depart significantly from that of $\alpha_0$. As shown in Fig.\,\ref{fig:convcooling}b, where $\alpha(\tilde{T}_{\rm g},\tilde{T}_0)$ is obtained by numerical integration of Eq.(\ref{eq:F}) for different values of $\tilde{T}_0$, the function $\alpha$ largely follows the $\alpha_0$ curve and only shifts slightly from $\alpha_0$ with increasing $\tilde T_0$. Further, the range of operating hot electron temperatures is now smaller, $\tilde{T_{\rm g}}\geq \tilde{T_0}$, as illustrated in Fig.\,\ref{fig:convcooling}b by curves that now start at $\tilde{T}_{\rm g} = \tilde{T}_0$. Although $\alpha$ is finite at $\tilde{T}_{\rm g}=\tilde{T}_0$, there is no net current at equal temperatures (as indicated by the empty circles at the beginning of the curves in Fig.\,\ref{fig:convcooling}b); a non-vanishing $\alpha$ at $\tilde{T}_{\rm g} = \tilde{T}_0$ arises from the differential ratio that characterizes the responsivity ${\cal R}$.

Considering losses, Eq.(\ref{eq:currentadim}) yields ${\cal J}_{\bf q}^{\perp}$ that can be sizable (see blue curves in Fig.\,\ref{fig:convcooling}c). In plotting Fig.\,\ref{fig:convcooling}c, we numerically integrated  Eq.(\ref{eq:currentadim}) and used $\phi=100\,{\rm meV}$ and $G_0=10\,{\rm mS}/\mu{\rm m^2}$ (see below for $G_0$ estimates). Further, we find that ${\cal J}_{\bf q}^{\perp}$ compares favorably with intrinsic electron-lattice cooling in graphene: (i) single-acoustic phonon cooling for pristine graphene ${\cal J}_{\bf q}^{\rm ac} $ (green curves), (ii) optical phonon cooling ${\cal J}_{\bf q}^{\rm op} $ (magenta curves), and (iii) disorder-assisted cooling ${\cal J}_{\bf q}^{\rm dis}$ (red curves), where we consider the degenerate limit ($\mu \gg T_{\rm g}$) for all cases \cite{macdonald,wong,song}:
\be
\begin{array}{l}
{\cal J}_{\bf q}^{\rm ac} = \gamma_{\rm ac} (T_{\rm g} - T_{0}),\quad {\cal J}_{\bf q}^{\rm dis} = \gamma_{\rm dis} ( T_{\rm g}^3 - T_0^3), \\ \\ {\cal J}_{\bf q}^{\rm op} = \gamma_{\rm op} \left[ N(\hbar\omega_{\rm op}/k_{\rm B}T_{\rm g}) - N(\hbar\omega_{\rm op}/k_{\rm B}T_{\rm 0})\right].
\end{array}
\label{eq:elatticecooling}
\ee
Here the prefactors are $\gamma_{\rm ac} = \hbar D^2 \mu^4 k_{\rm B}/ 8\pi \rho (\hbar v_{\rm F})^6$, $\gamma_{\rm dis} = 2D^2\mu^2k_{\rm B}^{3} / \rho c^2 \hbar (\hbar v_{\rm F})^4 k _{\rm F} \ell$ 
and $\gamma_{\rm op} = \hbar^2\omega_{\rm op}^3/24 \pi \rho a^4 v_{\rm F}^2$, with $D$ the deformation potential, $\rho$ the graphene mass density, $k_{\rm F}\ell$ the dimensionless disorder parameter, 
$a$ the lattice constant, $\omega_{\rm op}$ the optical phonon frequency, and $N(x)$ the Bose distribution. For the cooling mechanisms, we used $\mu = 100\,{\rm meV}$, $D = 20\,{\rm eV}$, $\rho = 7.6 \cdot 10^{-7}\,{\rm kg/m^2}$, $k_{\rm F}\ell = 50$, $a = 1.4\,{\rm \AA}$ and $\hbar \omega_{\rm op} = 0.2\,{\rm eV}$. Indeed, ${\cal J}_{\bf q}^\perp$ overwhelms both ${\cal J}_{\bf q}^{\rm ac}$ and ${\cal J}_{\bf q}^{\rm op}$, and is competitive with ${\cal J}_{\bf q}^{\rm dis}$, as shown in Fig.\,\ref{fig:convcooling}c. 

The hot carrier thermionic cooling channel, ${\cal J}_{\bf q}^\perp$, and the strong charge-energy current coupling it produces [Eq.(\ref{eq:ideal})], can manifest itself in large and non-monotonic responsivities in g/X photodetectors (Fig.\,\ref{fig:sens2}a). Accounting for energy balance in Eq.(\ref{eq:heatbalance}) we find a net responsivity given by
\be
{\cal R} = \frac{{\cal R}_0}{1+ \zeta}, \quad \zeta = \frac{{\cal J}_{\bf q}^{\rm loss}}{{\cal J}_{\bf q}^{\perp}},
\label{eq:netS}
\ee
where $\zeta$ quantifies losses. To estimate $\zeta$ for actual devices, we consider the disorder-assisted cooling power in graphene \cite{song}, ${\cal J}_{\bf q}^{\rm loss}\approx {\cal J}_{\bf q}^{\rm dis} = \gamma_{\rm dis} ( T_{\rm g}^3 - T_0^3)$ as an illustrative example, see Eq.(\ref{eq:elatticecooling}). Adopting the same procedure as described above, we find
\be
\zeta(\tilde{T}_{\rm g},\tilde{T}_0) = \zeta_* \times  \frac{\tilde{T}_{\rm g}^3 - \tilde{T}_0^3}{ \int_1^\infty  dx x \Delta f } \, \, , \, \, \zeta_* = \frac{e^2\gamma_{\rm dis}\phi}{G_0},
\label{eq:kappa}
\ee 
where the characteristic $\zeta$ is set by $e^2 \gamma_{\rm dis} \phi/G_0$. As expected, increasing the prefactor $\gamma_{\rm dis}$, for ${\cal J}_{\bf q}^{\rm dis}$, increases the losses to phonon scattering embodied in $\zeta$. Alternatively, increasing the conductance across the g/X interface enhances the thermionic channel.

In calculating ${\cal R}$ in Eq.(\ref{eq:netS}), we use the same parameter values as in Fig.\,\ref{fig:convcooling}: $\mu = \phi = 100 \, {\rm meV}$ and $k_{\rm F} \ell = 50$. The value of $G_0$ can be estimated from conductance measured in the dark state, $G_{\rm D}$, obtained in actual g/X devices at equilibrium $T_{\rm g}=T_0$ (for example g/Si Schottky junctions in Refs.\,\cite{gschottky1,gschottky2}). Indeed, under an infinitesimally small potential bias $\delta V_{\rm b}$, we can approximate $\Delta f$ in Eq.(\ref{eq:F}) as $\Delta f(x) =[e^{x}/(e^{x}+1)^2] \times e\delta V_{\rm b}/T_{0}$ due to the small chemical potential difference $e\delta V_{\rm b}$ between G and X. Integrating over $x$ in Eq.(\ref{eq:F}), we obtain $G_{\rm D}=G_0/(1+e^{\phi/T_0})$. In a typical scenario $T_0 \ll \phi$, the conductance in the dark state is exponentially suppressed with increasing temperature as $G_{\rm D} \approx G_0 \,{\rm exp}(-\phi/T_0)$, in agreement with the qualitative behavior observed in Refs.\,\cite{gschottky1,gschottky2,gschottky3,gsolar,britnellscience,ws2heterostructures,britnellnanolett}. To give an estimate of the range of conductances achievable in g/X devices, $G_{\rm D}$ in these experiments report $G_{\rm D} \sim 0.1 - {\rm several} \times \mu{\rm S}/\mu{\rm m}^{2}$ for $\phi_{\rm Si}\sim 0.3\,{\rm eV}$ with $T_0$ at room temperature. This gives $G_0$ in the $1-100\,{\rm mS}/\mu{\rm m^2}$ ballpark [for ${\cal R}$ in Fig.\,\ref{fig:sens2}c, we used $G_0 = 2,10,50\,{\rm mS}/\mu{\rm m^2}$, which correspond to $\zeta_* = 5, 1, 0.2$, see Eq.(\ref{eq:kappa})].

As shown in Fig.\,\ref{fig:sens2}b, $\zeta$ exhibits a clear non-monotonic dependence on $T_{\rm g}$ characterized by two regimes: (i) small $\tilde{T}_{\rm g}\ll 1$,  ${\cal J}_{\bf q}^{\perp}$ is exponentially suppressed by the transport barrier $\phi$, thus ${\cal J}_{\bf q}^{\rm dis}$ dominates, (ii) large $\tilde{T}_{\rm g} \gg 1$, we find that ${\cal J}_{\bf q}^{\perp}$ scales as $T_{\rm g}^2$, and rises less steeply than the $T_{\rm g}^3$ power law of supercollision cooling. Hence, there is a ``sweet spot'' for observing a competitive thermionic channel ${\cal J}_{\bf q}^{\perp}$. The optimal value occurs for temperatures $T_{\rm g}/\phi \approx 0.5$ (see Fig.\,\ref{fig:sens2}b), with minimum $\zeta_{\rm{min}}\approx 1.85 \times \zeta_*$. This can be estimated from Eq.(\ref{eq:kappa}) in the limit $\tilde{T}_{\rm g} \ll 1$ and $T_0=0$, where the above-mentioned optimal values are obtained from minimization of the equation $\zeta / \zeta_* \approx \tilde{T}_{\rm g}^2 e^{1/\tilde{T}_{\rm g}}$ (Fig.\,\ref{fig:sens2}b). 

The responsivity ${\cal R}$ in Eq.(\ref{eq:netS}) mirrors $\zeta$ to display a non-monotonic dependence on $T_{\rm g}$, peaking at a temperature $k_{\rm B}T_{\rm g} \approx \phi/2$, as shown in Fig.\,\ref{fig:sens2}c. Peak responsivities in the range ${\cal R} \approx 1-10\, {\rm A/W}$ are obtained within our model. Indeed, for large $G_0 = 50\,{\rm mS}/\mu{\rm m^2}$ (corresponding to $\zeta_* = 0.2$), ${\cal R}$ starts to approach the ideal case, ${\cal R}  = {\cal R}_0$ (dashed black line). The non-monotonic dependence of ${\cal R}$ as a function of $T_{\rm g}$ provides a clear fingerprint of the competition between thermionic energy transport and conventional electron-phonon cooling. Since the Schottky barrier heights can be tuned by the applied gate voltage, the peak temperature $k_{\rm B}T_{\rm g} \approx \phi/2$ is gate tunable. Further, the scaling of $\phi$ and the device conductance $G_0$ also provides experimental knobs with which to adjust the responsivity of the device.

The optimal responsivity ocurring at $k_{\rm B}T_{\rm g} \approx \phi/2$ is an important characteristic for the design of graphene photodetectors. Indeed, given that $T_{\rm g} < 2000\,{\rm K}$ in realistic situations, Schottky barriers in the $100\,{\rm meV}$ ballpark allow operation of the g/X photodetector near optimal responsivities (i.e. near minimum $\zeta$). These values of $\phi$ can be achieved, for instance, in graphene-WS$_2$ devices \cite{ws2heterostructures}. 

Although g/X photodetectors allow in situ  control of $\phi$ by electrostatic doping, it is important to note that several parameters of the model vary implicitly with $\phi$. On the one hand, changes in $\phi$ also induce changes in graphene doping, thus modifying the electronic cooling power. Further, when $\phi$ becomes smaller than the incoming photon energies, photo-emission of primary carriers over the barrier competes with thermalization by electron-electron interactions. In this case, a smaller amount of the incident power is captured in the hot-carrier distribution.

Naturally, there are other mechanisms for losses that affect the responsivity. For instance, lateral (in-plane) heat currents, ${\cal J}_{\bf q}^{\parallel} = -\nabla \cdot (\kappa_\parallel \nabla T_{\rm g})$, can transport heat towards the contacts in small devices. To estimate this effect, we use the Wiedemann-Franz relation, $\kappa_\parallel(T_{\rm g})=(\pi^2/3e^2) \times k_{\rm B}^2 T_{\rm g} \sigma$, where $\sigma$ is the in-plane electrical conductivity of graphene. For the relevant regime of moderate to high temperatures, $T_{\rm g}\gtrsim \phi$, we can approximate ${\cal J}_{\bf q}^{\perp}\approx G_0 k_{\rm B}^2 T_{\rm g}^2/e^2 = \gamma^{\perp}(T_{\rm g})\times T_{\rm g}$ [cf. Eq.(\ref{eq:current})]. As a result, we find a cooling length $\xi_\parallel = \sqrt{\kappa(T_{\rm g})/\gamma^{\perp}(T_{\rm g})}$ coming from the thermionic channel that is independent of $T_{\rm g}$. Using a uniform in-plane $\sigma \sim 1\,{\rm mS}$ \cite{novoselov}, we find $\xi_\parallel \approx 0.6 \, \mu{\rm m}$, so that vertical energy extraction dominates over in-plane thermal conduction for sufficiently large devices with size $L > \xi_{\parallel}$.

We note that interactions with the substrate can result in cooling via surface optical phonons. These losses will vary for different substrate (X) choices and are only significant when X is a polar material \cite{shytov}. Importantly, we do not expect them to be relevant in non-polar materials, e.g. X $=$ silicon. 

Lastly, it is interesting to note that g/X photodetectors can also operate at low photon energies, $\hbar\omega \leq 2\mu $. In this regime, conventional Drude absorption from ambient carriers directly captures incident radiation. This contrasts with conventional semiconductor photodetectors, that do not absorb light below the semiconductor bandgap. A tantalizing possibility is to use g/X Schottky junctions within the mid IR - THz bandwidth where presently-available technologies offer lackluster performance \cite{chan2007,rogalski}. 

In summary, graphene Schottky junctions host tunable interfaces across which energy transport can be engineered, exemplified by thermionic-dominated transport regime wherein energy and charge currents are strongly coupled. Fingerprints of the thermionic-dominated regime include high responsivities on the order of ${\cal R} \sim 1-10$ A/W, and a non-monotonic dependence of ${\cal R}$ on electron temperature (or pump power) in g/X photodetectors. The large degree of in situ tunability allows optimization of the g/X interface for different applications and irradiation conditions; vertical hot carrier convection opens up new vistas to efficiently harvest photon energies over a wide spectral range, utilizing the entire exposed graphene area as a photoactive region.

We are grateful to useful discussions with M. Baldo, M. Kats, L. Levitov. We also thank V. Fatemi, A. Frenzel, and K. Tielrooij for a critical reading. JFRN and MSD acknowledge financial support from the National Science Foundation Grant DMR-1004147. JCWS acknowledges support from a Burke Fellowship at Caltech. 

\vspace{3mm}

{\it JCWS Current Address:} Institute of High Performance Computing Singapore, and Division of Physics and Applied Physics, Nanyang Technological University.

\end{document}